\documentclass[12pt]{iopart}
\usepackage{iopams}
\jl{6}

\newtheorem{Proposition}{Proposition}[section]

\newtheorem{remark}[Proposition]{Remark} 
\def\R{\ensuremath{\mathcal{R}}}   
\def\ihat{\^{i}}
\def\jhat{\^{j}}
\def\khat{\^{k}}

\begin{document}
\title{The orientability of spacetime} 
\author{Mark J Hadley}

\address{Department of Physics, University of Warwick, Coventry
CV4~7AL, UK\\ email: Mark.Hadley@warwick.ac.uk}

\begin{abstract}

Contrary to established beliefs, spacetime may not be
time-orientable.  By considering an experimental test of time
orientability it is shown that a failure of time-orientability of a
spacetime region would be indistinguishable from a particle
antiparticle annihilation event.  
\end{abstract}

\submitted
\maketitle

\section{Introduction} \label{introduction}

It  is widely  believed that  spacetime  must be  both orientable  and
time-orientable \cite{visser}. Arguments are that there is no evidence
of a lack  of orientability and that a  non-orientable spacetime would
be incompatible with the observed violations of P (parity) and T (time
reversal) invariance. These arguments are shown to be false.

This  paper   aims  to    answer   the   questions:  {\em   How    can
time-orientability be  tested?}   and {\em  Is  a  non-time orientable
spacetime consistent with observations?}.  The focus of this paper  is
on experiments a physicist would have to do  to test the orientability
of a spacetime region. 

The work  is motivated by results suggesting  that non-time orientable
spacetime   regions  might   be   used  to   explain  quantum   theory
\cite{hadley98}, the existence of electric charge \cite{diemer_hadley}
and  spin half  \cite{hadley2000}.  Topological  models  of elementary
particles  within the  equations and  framework of  general relativity
require a breakdown of the causal structure to allow topology changing
interactions.   A  non-timeorientable  spacetime  structure  can  also
exhibit  net electric charge  from source  free Maxwell  equations and
would naturally have the transformation properties of a spinor under
rotations.

Mathematically, the orientability of a manifold is a simple and
unambiguous property. A few mathematical definitions and theorems are
given in the next section.  This paper looks at how the mathematical
definitions can be used in practice to answer questions about
spacetime.  Some simple experiments are postulated; they follow the
mathematical definitions very closely, but the interpretation remains
ambiguous.

\section{Preliminary results}

This paper is exclusively concerned with connected spacetimes.

\begin{remark}
Locally spacetime is always orientable.
\end{remark}
Spacetime is modelled by a 3+1 dimensional semi-Riemannian
manifold. Therefore mappings to flat Minkowski spacetime exist in a
region of every point. Minkowski space-time is orientable hence
spacetime is locally orientable. It follows that orientability is a
global rather than a local property.

\begin{remark}
On a non-orientable spacetime there exists a closed path along which a
consistent orientation cannot be defined. \label{rem:path}
\end{remark}
This follows from the definition of orientability and topological
arguments.

\begin{remark}If spacetime is not orientable then through every point
there exists a path on which an orientation cannot be defined.
\end{remark}
This follows simply from remark~\ref{rem:path} above and assumed
connectedness of spacetime. If A has a non-orientable path, C, through
it and X is any other point. Then by connectedness there exists a
path, P, from X to A and a path, P', from A to X. A non-timeorientable
closed path through X is constructed by joining the paths P-C-P'.

Putting the three remarks together, spacetime could be orientable for
all paths except those going through a small remote region.  For all
normal purposes such a spacetime would appear orientable. Only an
experiment that probed the remote region \R\ could detect the
non-orientability. However, Non-orientability in a connected spacetime
cannot be strictly localised as there would be a curve through every
spacetime point that went through the non-timeorientable region.

Mathematically, orientability is a global property of spacetime. By
analogy with a m\"obius strip it is possible that paths comparable in
size to the Universe could be non-orientable. However a non-orientable
region could be microscopic in size: for example a non-orientable
wormhole with dimensions comparable to the Plank length. This work is
motivated by the possibility that elementary particles may be regions
of non-trivial topology, but the arguments are generally
applicable.

It could of course be possible for all {\em timelike} paths in a
non-orientable spacetime to be orientable. A non-orientable region
behind an event horizon would be such an example.

The observed violation of T-invariance implies that an observer can
unambiguously define the direction of time. This is compatible with a
localised time-reversing region, because the boundary conditions (the
expanding Universe) define a time direction and because a consistent
time direction could still be defined if the experimenter never probed
the time-reversing region. Microscopic localised nonorientable regions
as described in \cite{diemer_hadley} are therefore consistent with
observations.

\section{Tests of space and time orientability}

This section examines classical tests of the orientability of
spacetime - they are based upon classical general relativity and
classical physics. It cannot be overemphasised that this argument is
not based on quantum field theory or even quantum mechanics, it is
classical - classical mechanics and classical general relativity.
 
Consider a space that is flat and space-orientable everywhere except
for a region \R. In spacetime \R\ would be a tube. The construction of
spacetimes such as this, with all permutations of orientability, are
described in \cite{diemer_hadley}: Sorkin \cite{sorkin} constructs a
non-orientable wormhole, which is also asymptotically flat.  Gibbons
and Herdeiro \cite{gibbons} give an asymptotically flat example for a
supersymmetric rotating black hole. The arguments in this paper are
independent of the exact form of the metric in \R.

The orientation of \R\ is tested by sending a probe into \R. The probe
contains a triad of unit vectors \ihat, \jhat and \khat.  The triad
defines a space-orientation continuously along the path taken by the
probe. The path (timeline) of the probe and the path (timeline) of the
observer together define a closed loop. An orientation can be defined
in the laboratory, excluding region \R. If the probe that emerges from
\R\ has a different orientation from that defined in the rest of the
laboratory then space is not orientable (see Figure~\ref{fig:space}).

An example of a region \R\ that is not time-orientable is easily
constructed (see for example Ref.~\cite{hadley2000}): remove a ball
from space - that is a world-tube from spacetime.  Then identify
opposite points of the sphere using the antipodal map.  At the same
time identify $t$ with $-t$ in a similar manor to the creation of a
M\"obius strip.  Like a M\"obius strip this example is continuous and
flat.

A similar test of the time orientability of \R\ would need a probe
that defined a time direction continuously along a trajectory through
\R\ and back to the observer.  A clock would be a suitable probe, with
the positive time direction defined by increasing times displayed on
the clock.  However the experiment is not so simple. Consider a clock
that enters \R\ at an Observer time $\tau$ and emerges at a later
observer time but counting backwards (see Figure~\ref{fig:time1}).
This is not a demonstration of non-time orientability, because in this
experiment, the clock increases in value and then decreases. At some
point in the path it attains a maximum reading and at that point it
does not define a time direction.

In a true test of time orientability on a region \R\ through which
time cannot be oriented, the clock readings would increase steadily
along the path taken by the clock.  Before entering \R\ the observer
sees the time values increasing on the clock.  When the clock exits at
a time $\rho$ the increasing clock times would be at ever decreasing
values for the observer time. The observer would still see a backwards
counting clock, but only at times before $\rho$ (see
Figure~\ref{fig:time2}).

This might seem like a clear example of a successful demonstration of
a failure of time orientability.  The author would be sympathetic to
such a view!  However the observer sees the backwards (clock-time
counting backwards) moving clock entering \R\ as well as the forward
counting clock entering \R. At observer times greater than $\rho$ the
observer sees no clock at all - neither inward nor outward moving. So
the observer can interpret the experiment as a clock and an anti-clock
entering \R\ and annihilating each other.

It must be emphasised that this is still a description in terms of
classical physics. The expression {\em anti-clock} is a logical one -
it does not refer to anti-particles in the normal sense - indeed it is
totally independent of the construction of the clock. If you expect
the number of clocks to be preserved, then describing the backward
counting clock as an anti-clock achieves the objective.  Using the
concept of an anti-clock, the experiment has zero clocks most of the
time. Before $\tau$ there is a clock and anti-clock after $\rho$ there
are no clocks.  The alternative is to postulate two rather different
real objects (the clock and the backward counting clock) that both
enter \R\ and vanish. The analogy shares with anti-particles the fact
that both the clock and anti-clock must exist for this to happen.
Unlike the conventional particle -antiparticle annihilation there does
not appear to be any conservation of energy.

Gibbons and Herdeiro make similar observations for a particle
traversing their supersymmetric rotating black hole.

In general $\tau$ and $\rho$ would be different.  In a normal
particle-antiparticle annihilation event they would be equal.  The
difference between $\tau$ and $\rho$ depends upon the geometry. For a
similar construction to that described above the identification of
$\tau -t$ with $\tau+t$ (rather than $t$ and $-t$) would result in
$\tau$ and $\rho$ being equal. A simple would be just a special case
of this

To make matters even worse, the experiment is dependent upon the
observer. Since any attempt to stop the backward counting clock from
entering \R\ would be inconsistent with the clock entering and leaving
a time reversing region. So not only is the anti-clock description a
possible alternative description, the time reversing event {\em
requires} the existence of anti-clocks (time reversed clocks) to exist
before the experiment takes place.  Since a clock is normally a
macroscopic object composed of atoms, anticlocks do not exist in the
sense of clocks composed of antiparticles. Therefore, an apparently
simple test may be physically impossible to carry out.

This highlights a weakness in the way the problem was posed.  Space
orientability is a property of space.  However timeorientability can
only be a property of spacetime.  Furthermore it is a global property
of spacetime.  It would appear that the existence or otherwise of a
time reversing region is dependent upon the observer.  A probe and an
anti-probe must both be prepared for a consistent result, in which
case it could be argued that the existence of the time reversing
region depends upon the experiment being performed to test it.

\section{Quantum field theory}
Although the preceding argument referred to anti-clocks, this was a
purely logical classical description.  Quantum Field theory, which
correctly describes both particles and antiparticles - and indeed
requires anti-particles for completeness, was not being used.  The
similarities of these classical arguments to some quantum phenomena is
not coincidental.  It has already been shown~\cite{hadley97} that
quantum logic can be derived from classical general relativity with
acausal spacetimes.

It is not possible to define a spinor field on a non-time orientable
spacetime \cite{geroch}.  Since fermions exist, this well-known result
has been cited as evidence that time must be orientable. However the
argument relies on a realist interpretation of the wavefunction and
the false assumption that a wavefunction is defined at each spacetime
point.  In fact a wavefunction is a function defined on a
3N-dimensional configuration space where, N, is the number of
particles. See Ballentine \cite{ballentine} Chapter 4 for the clearest
discussion of arguments against a realist interpretation of a
wavefunction (also \cite{ballentine70}).

Fermion fields are used to calculate the probabilities of results of
experiments.  The extension of a fermion field to inaccessible
microscopic non-orientable regions is neither relevant nor
particularly meaningful, since measurements cannot take place within
these regions.  While on a spacetime with accessible time-reversing
regions, the inability to define a spinor field, simply means that QFT
cannot be used to calculate probabilities of the results of
experiments on a space time that is not timeorientable a conclusion
that is hardly surprising!

\section{Conclusion}

Although  the analysis  is clear,  the interpretation  is by  no means
obvious.  The  following  statements  can  all  be  supported  by  the
arguments above:
\begin{enumerate}
\item  Spacetime   is  not  time   orientable.  Particle  antiparticle
annihilation events are evidence of this.
\item  A  failure  of  time orientability  and  particle  antiparticle
annihilation are indistinguishable.  They are alternative descriptions
of the same phenomena.
\item Time orientability is untestable.
\item  Non  time orientability  cannot  be  an  objective property  of
spacetime  because the  outcome  of  our test  would  depend upon  the
observer.
\end{enumerate}

The conventional  view of the experiment  in Figure~\ref{fig:time2} is
that  it depicts a  particle antiparticle  annihilation. This  view has
developed  and  now predominates  for  historical  reasons.  It is  an
interpretation  that conforms  to  the classical  paradigm -  modelling
Nature in  terms of initial conditions, unique  evolution and observer
independent outcomes.  The classical paradigm  is clearly inconsistent
with a failure of time-orientation.

The explanatory power of a non-timeorientable spacetime in relation to
interactions  in  geon models,  electric  charge  and spin-half,  adds
weight to the interpretation of Figure~\ref{fig:time2} as a demonstration
of the non-timeorientability of spacetime.

\setlength{\unitlength}{1mm} \input epsf
\begin{figure}[p]
\begin{picture}(150,200)
\put(15,-100){\makebox{\epsfbox{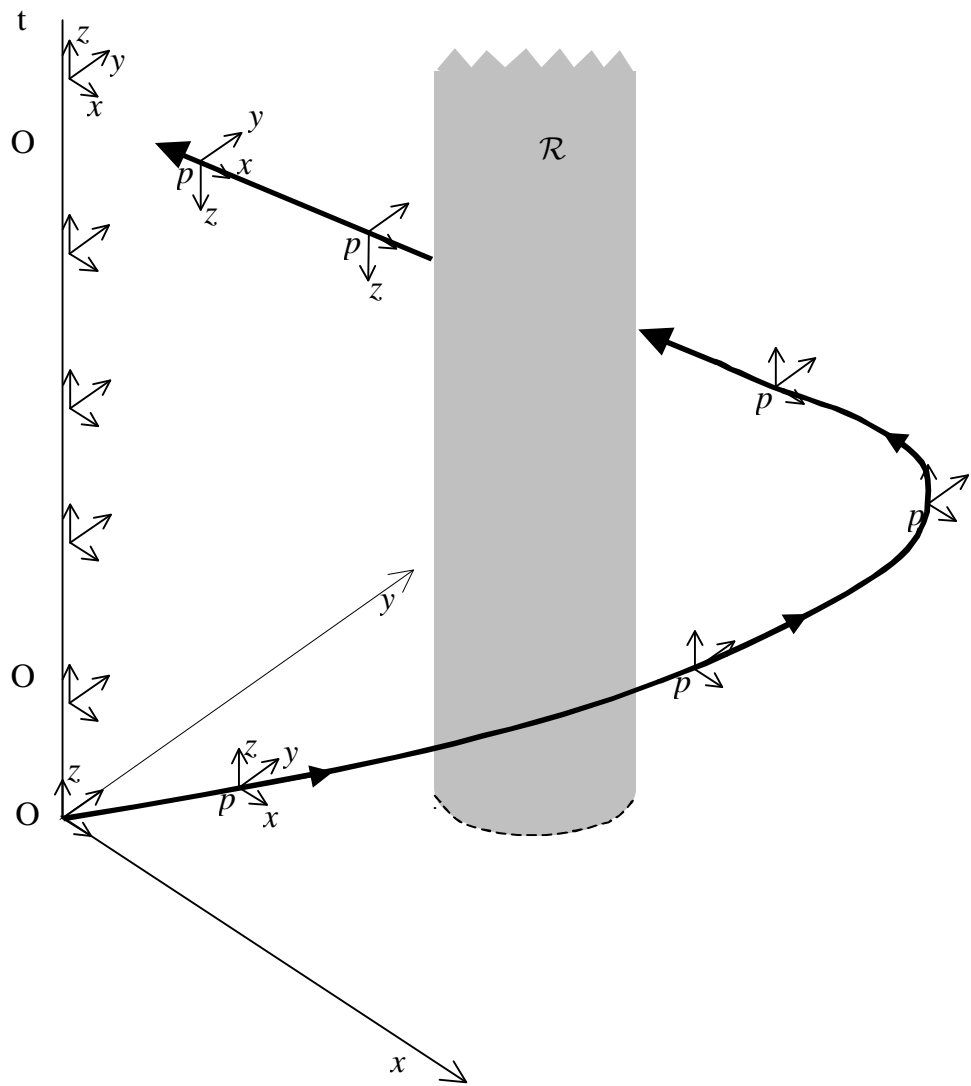}}   }
\put(98,125){\makebox{\R} }
\end{picture}
\caption{A spacetime  diagram showing a positive test  that region \R\
is not timeorientable.  Observer, O,  sends a probe, P, through \R, it
exits \R\ and returns to O with opposite orientation.}
\label{fig:space}
\end{figure}

\begin{figure}[p]
\begin{picture}(150,200)
\put(15,-100){\makebox{\epsfbox{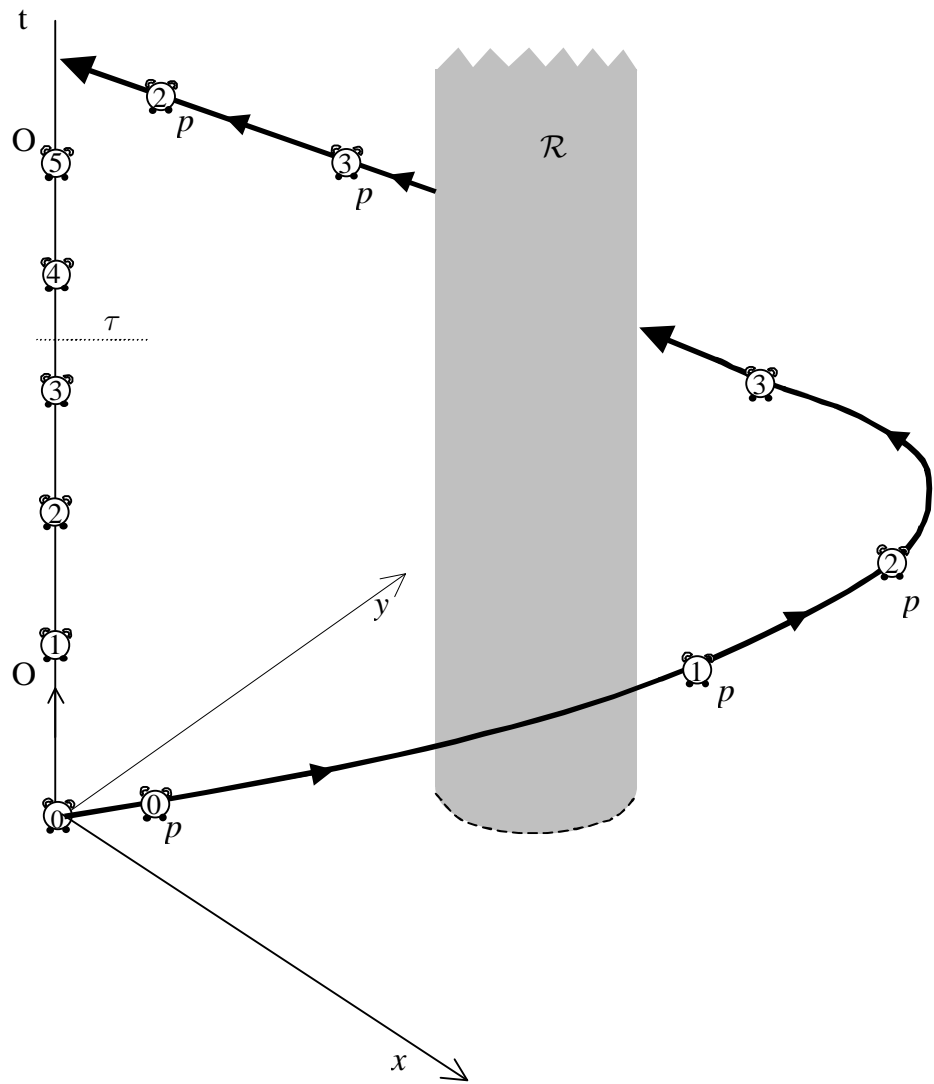}}    }
\put(40,102){\makebox{$\tau$} }
\put(98,125){\makebox{\R} }
\end{picture}
\caption{A flawed attempt to  test time-orientability: An Observer, O,
sends a clock,  P, through \R, it exits \R\ and  reappears counting
backwards.  This  is not  successful because the  clock has  failed to
define a time direction throughout the experiment.}
\label{fig:time1}
\end{figure}

\begin{figure}[p]
\begin{picture}(150,200)
\put(15,-100){\makebox{\epsfbox{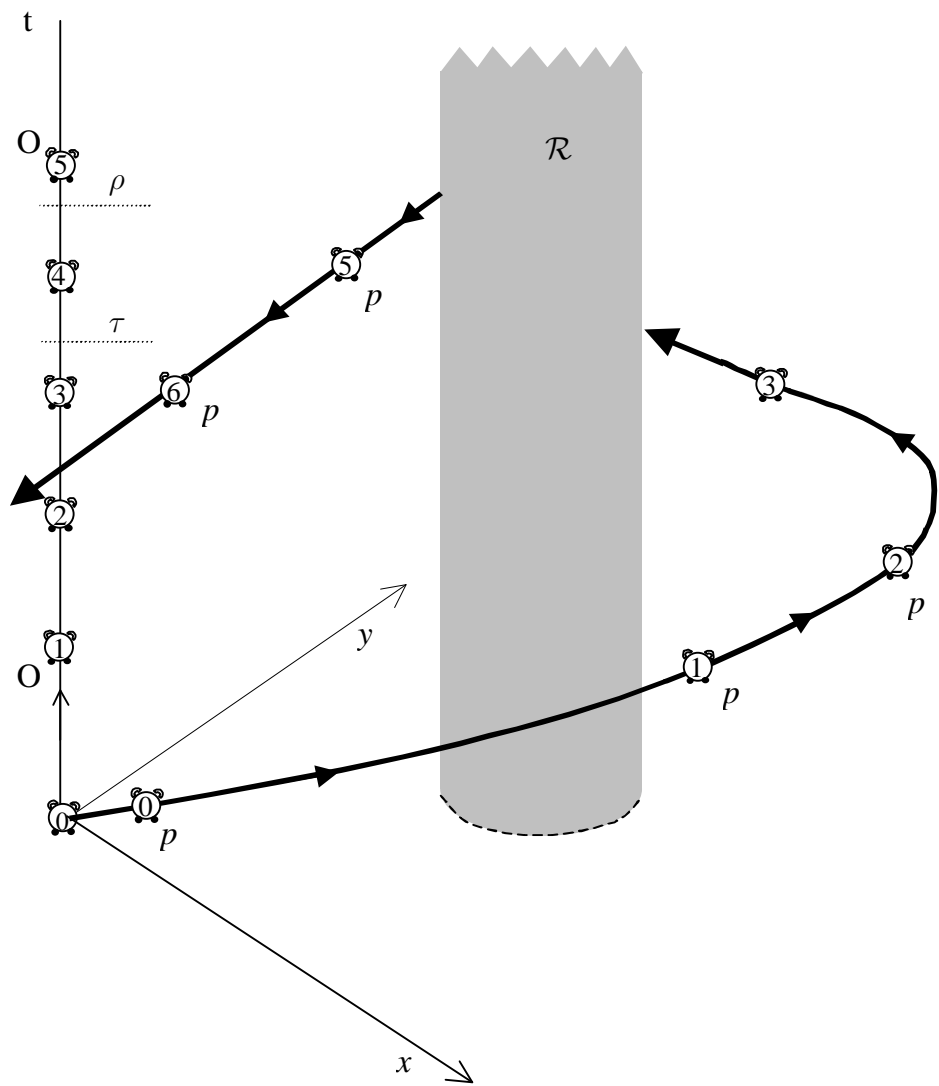}}      }
\put(98,125){\makebox{\R} } 
\put(40,121){\makebox{$\rho$} }
\put(40,102){\makebox{$\tau$} }
\end{picture}
\caption{The clock,  P, counts forward continually, it  defines a time
direction  throughout  its path.  The  observer,  O,  sees a  backward
counting clock  and a  forward counting clock  enter \R.  The observer
sees no clocks after observer time $\rho$.}
\label{fig:time2}
\end{figure}

\end{document}